\documentclass[preprint,aps,prc,showpacs,nofootinbib]{revtex4}
\usepackage{epsfig}
\usepackage{graphicx,amsmath}

\newcommand\nn{\nonumber}

\def\phi{\varphi}

\newcommand\ba{\begin{eqnarray}}
\newcommand\ea{\end{eqnarray}}

\begin{document}

\title{Radiative corrections to the cross section of $e^-+p\to \nu+n$ and the crossed processes.}

\author{E.~A.~Kuraev}
\affiliation{\it Bogoliubov Laboratory of Theoretical Physic, JINR,
141980 Dubna, Moscow region, Russia}

\author{S. Bakmaev}
\affiliation{\it Bogoliubov Laboratory of Theoretical Physic, JINR,
141980 Dubna, Moscow region, Russia}

\author{Yu.~M.~Bystritskiy}
\affiliation{\it Bogoliubov Laboratory of Theoretical Physic, JINR,
141980 Dubna, Moscow region, Russia}

\author{T.~V.~Shishkina}
\affiliation{\it Belarus State University, Dept. of Physics, 220050
Minsk , Belarus}

\author{O.~P.~Solovtsova}
\affiliation{\it Bogoliubov Laboratory of Theoretical Physic, JINR,
141980 Dubna, Moscow region, Russia}
\date{\today}
\begin{abstract}
   Born cross section and the radiative corrections to its  lowest order are considered
   in the frame work of QED with structureless nucleons including the emission of virtual and
   real photons. Result is generalized to take into account radiative corrections
   in higher orders of perturbation theory in the leading and next-to leading logarithmic
   approximation. Crossing processes are considered in the leading approximation.
\end{abstract}
\maketitle

\section{Introduction}

The process of neutron $\beta$ decay is  rather completely
investigated experimentally. Its theoretical description has not
been sufficiently studied up to now, due to uncertainties connected
with effects of strong interactions which arise of calculations of
the radiative corrections (RC). Research into neutrino-deuterium
inelastic scattering was carried out in experimental set-ups like
the Sudbury Neutrino Observatory (SNO)\cite{Towner}. The increasing
accuracy of experiments and the importance of adequate theoretical
description of the neutrino sector of Standard Model (SM) is the
motivation of our paper. It is devoted to calculation of RC in the
framework of QED with the point-like nucleons and omission the
electro-weak effects. The last ones were considered in a series of
papers (see \cite{Hollik} and the references therein). Below we
consider the process \ba e(p_e)+P(p) \to \nu(p_\nu)+N(p_n), \ea and
the crossed processes \ba
\nu(p_\nu)+N(p_n) \to e(p_e)+P(p),  (\nu n), \nonumber \\
\bar{\nu}(p_\nu)+P(p) \to e^+(p_e)+N(p_n),   (\nu p). \ea Our work
is performed in terms of the old formulation of weak processes which
is valid for the intermediate region of energies of electron $E$ (we
suppose proton to be at rest -Laboratory Frame) \ba m=m_e\ll E\ll
M_Z. \ea

\section{Born cross section. Kinematics}

The matrix element of the process $e p \to \nu n$ in the Born
approximation has the form: \ba
M=\frac{G}{\sqrt{2}}V_{ud}\bar{u}_\nu(p_\nu)\gamma_{\alpha}(1+\gamma_5)u(p_e)\bar{u_n}\gamma_{\alpha}(1+\rho\gamma_5)u(p),
\ea with $G,V_{ud}$ being the Fermi constant, the
Cabbibo-Kobayashi-Maskawa quark-mixing matrix element and
$\rho=g_A/g_V=1.25$ being the ratio of axial and vector coupling
constants.

The cross-section has the form \ba
d\sigma=\frac{1}{8s}\sum|M_{if}|^2
d\Gamma, \ea with \ba \sum|M_{if}|^2=8G^2|V_{ud}|^2 S_0(p_e,p),\nn\\
S_0(p_e,p)=s^2(1+\rho)^2+u^2(1-\rho)^2-2M^2(s+u)(1-\rho^2); \ea
where \ba s=2ME;\quad u=-2p_ep_n=-2E(E_n-P_n c),\quad
p_e^2=m^2,\quad p^2=p_n^2=M^2, \nn \ea and the phase volume \ba
d\Gamma=\frac{d^3p_\nu}{2E_\nu}\frac{d^3p_n}{2E_n}(2\pi)^{-2}\delta^4(p_e+p-p_\nu-p_n).
\ea

Performing the integration on neutrino momentum we write the phase
volume as \ba d\Gamma=\frac{c d
c}{2\pi}\frac{ME(M+E)^2}{[(M+E)^2-E^2c^2]^2}, \ea where $E$- is the
energy of initial electron, $c=\cos\theta$ and $\theta$ is the angle
in Laboratory Frame between the 3-momenta of electron and neutron.
The energy $E_n$ and the value of 3-momentum of neutron $P_n$ are:
\ba E_n=M\frac{(E+M)^2+E^2c^2}{(E+M)^2-E^2c^2};\quad
P_n=M\frac{2E(E+M)c}{(E+M)^2-E^2c^2}. \ea Another form of phase
volume can be used for investigating the neutron energy distribution
\ba d\Gamma=\frac{d E_n}{8\pi E}. \ea

            \section{Virtual photon emission and contribution counter term}

Consider now the one-loop virtual radiative correction: the doubled
interference of Born amplitude with one with the emission of virtual
photon taken into account. As in Born amplitude we suggest that the
loop momentum as well as the external ones are much small compared
to the $W$-boson mass. Only one triangle Feynman diagram is
relevant.  The corresponding contribution to the summed up spin
states of the square of matrix element is
 \ba
 \Delta\sum|M|^2_{virt}=(G_FV_{ud}/\sqrt{2})^2\frac{8\alpha}{\pi}\int\frac{d^4k}{i\pi^2}\frac{T}{(0)(1)(2)},\nn\\
 (0)=k^2-\lambda^2);\quad (1)=(p_--k)^2-m^2+i0;\quad  (2)=(p+k)^2-M^2+i0,
 \ea
where $k$ is the loop 4-momentum, $\lambda$ is the fictitious photon
mass, and \ba
T=\frac{1}{4}Tr \hat{p}_\nu\gamma_\alpha(1+\gamma_5)(\hat{p}_e-\hat{k})\gamma_\mu\hat{p}_e\gamma_\eta(1+\gamma_5) \nn \\
\frac{1}{4}Tr
(\hat{p}_n+M)\gamma_\alpha(1+\rho\gamma_5)(\hat{p}+\hat{k}+M)\gamma_\mu(\hat{p}+M)\gamma_\eta(1+\rho\gamma_5).
\ea The next step is to perform the loop momentum integration. To
avoid
 ultraviolet divergences we must to introduce the cut off parameter
$|k^2|<\Lambda^2$. It is natural to choose the cut-off parameter to
be equal to the $W$-boson mass:
$$
\Lambda=M_W.
$$
Using the Feymnan trick to join the denominators we obtain the
expression \ba \int\frac{d^4k}{i\pi^2}\int\limits_0^1
dx\int\limits_0^12y d y\frac{T}{[(k-yp_x)^2-D]^3}, \nn \ea with
$D=y^2p_x^2+(1-x)\lambda^2+i0,p_x^2=x^2m^2+(1-x)^2M^2-sx(1-x)$.

The computation of the traces leads to: \ba
T=A_0+A_e(2kp_e)+A_n(2kp_n)+A_p(2kp)+A_{en}(2kp_e)(2kp_n)+\nn
\\A_{ep}(2kp_e)(2k p)+A_{pn}(2k p)(2k p_n)+ A_{ee}(2k
p_e)^2+A_{nn}(2k p_n)^2+A_{kk}k^2, \ea with \ba A_0=2s S_0;\quad
A_e=2(s+M^2)[s(1+\rho)^2-u(1-\rho)^2]; \nn \\
A_n=2s[u(1-\rho)^2-M^2(1-\rho^2)]; \nn \\
A_p=-2s^2(1+\rho)^2+2(2s+u)M^2(1-\rho^2); \nn \\
A_{np}=u(1-\rho)^2; A_{ee}=(s+u)(1-\rho)^2-4M^2\rho(1-\rho); \nn \\
A_{kk}=-2u(u+s)(1-\rho)^2-4s^2(1+\rho)^2+4M^2(s+u)(1+\rho-2\rho^2); \nn \\
A_{ep}=u(1-\rho)^2-4M^2(1-\rho^2); \nn \\
A_{en}=-(2s+u)(1-\rho)^2+4M^2\rho(1-\rho). \ea Further integration
(we are interested only in the real part) gives \ba
[i;i_x;i_{\bar{x}b};i_{x\bar{x}};i_{x\bar{x}b};i_{\bar{x}b\bar{x}b}]=\int\limits_0^1\frac{dx}{p_x^2}[1;x,1-x;x^2;x(1-x);(1-x)^2]= \nn \\
-\frac{1}{s}L;-\frac{1}{s}L+\frac{1}{a}L_M;-\frac{1}{a}L_M; \nn \\
-\frac{1}{s}L+\frac{1}{a}+\frac{s+2M^2}{a^2}L_M;-\frac{1}{a}-\frac{M^2}{a^2}L_M; \nn \\
\frac{1}{a}-\frac{s}{a^2}L_M, \ea \ba i_p=\int\limits_0^1 dx
\ln\frac{p_x^2}{M^2}=-2+\frac{s}{a}L_M, \ea with
$L=\ln\frac{4E^2}{m^2}$ being the so-called "large logarithm",
$L_M=\ln\frac{2E}{M}$ and $a=s+M^2$. Finally, \ba
 I=\int\frac{d^4k}{i\pi^2}\frac{1}{(0)(1)(2)}=-\int^1_0\int^1_0\frac{dx y dy}{D}=-\int^1_0\frac{dx\ln\frac{p^2_x}{\lambda^2}}{2p^2_x}. \nn\\
 \ea
 Performing the integration on Feynman parameters we obtain
 \ba
 Re I=\frac{1}{2s}[\frac{1}{2}L^2+2L\ln\frac{m}{\lambda}-L^2_M-\pi^2-2Li_2(1+\frac{M^2}{s})],
\ea with the Euler dilogarithm defined as \ba
Li_2(z)=-\int\limits_0^z\frac{d x}{x}\ln(1-x). \ea The result is:
\ba 2\sum
M_B^*M_{virt}=\nn\\-\frac{\alpha}{\pi}S_0[-L+L\ln\frac{m}{\lambda}+\frac{1}{4}L^2-\frac{1}{2}L_m^2-\frac{\pi^2}{2}-
Li_2(1+\frac{M^2}{s})-A L_\Lambda + K_V], \ea where $L_\Lambda =
\ln(M_W^2/M^2)$ and \ba
    A = \frac{1}{2S_0}\left[4s^2(1+\rho)^2+u^2(1-\rho)^2-5M^2(s+u)(1-\rho^2)\right].
\ea

A rather complicated expression for $K_V$ is  given in Appendix.

Since we work within the unrenormalized theory, we must take into
account the fermion (of mass $m$) wave function renormalization: \ba
Z=1-\frac{\alpha}{2\pi}[\frac{1}{2}\ln\frac{\Lambda^2}{m^2}+\ln\frac{\lambda^2}{m^2}+\frac{9}{4}].
\ea

Keeping in mind to apply this procedure for both electron and proton
we obtain the contribution to the summed on spin states of the
matrix element squared \ba
\Delta|M|^2_c=-\frac{\alpha}{2\pi}S_0[L_\Lambda+\frac{9}{2}-\frac{1}{2}L+L_M].
\ea

                 \section{Real soft and hard photon emission}

A standard way to take into account the emission of additional soft
(in the Laboratory frame) real photon consists in calculation of the
3-dimensional integral \ba
\frac{d\sigma_{soft}}{d\sigma_B}=-\frac{4\pi\alpha}{16\pi^3}\int\frac{d^3k_1}{\omega_1}
(\frac{p_e}{p_ek_1}-\frac{p}{pk_1})^2|_{\omega_1<\Delta E},\quad
\Delta E\ll E. \ea The standard calculation leads to \cite{Hooft}:
\ba d\sigma_{soft}=d\sigma_B\frac{\alpha}{\pi}[(L-1)\ln\frac{\Delta
E}{E}+(L-2)\ln\frac{m}{\lambda}+ \frac{1}{4}L^2+1-\xi_2],\quad
\xi=\frac{\pi^2}{3}. \ea The total sum including the Born cross
section and corrections arising from the wave function
renormalization, and ones arising from taking into account emission
of virtual and soft real photons is free from infrared singularities
\ba
\frac{d\sigma_B+d\sigma^{s+v+c}}{d\sigma_B}=1+\frac{\alpha}{2\pi}[P_\Delta(L-1)+\frac{\alpha}{\pi}K_{svc}-
 \frac{\alpha}{\pi} A L_\Lambda]-\frac{\alpha}{\pi}\ln\frac{\Delta E}{E},
\ea with \ba P_{\Delta}=2\ln\frac{\Delta E}{E}+\frac{3}{2}. \ea We
note that the term $(L-1)P_\Delta$ containing the "large logarithm"
can be associated with the kernel of the Altarelly-Lipatov-Parisi
evolution equation  of twist-2 operators \cite{Altarelly}.

Results obtained here for virtual and soft real photons emission are
in agreement with ones obtained in \cite{Yokoo:1973, Yokoo76}

The emission of the real hard photon matrix element has the form \ba
M^\gamma=\frac{G}{\sqrt{2}}V_{ud}{\bar{u}_\nu}(p_\nu)\gamma_\alpha(1+\gamma_5)[\frac{\hat{p}_e-\hat{k}+m}{-2p_e
k}\hat{e}] u(p_e)
\cdot\bar{u}(p_n)\gamma_\alpha(1+\rho\gamma_5)u(p)\nn\\-\bar{u}_\nu(p_\nu)\gamma_\alpha(1+\gamma_5)u(p_e)\cdot\
\bar{u}(p_n)\gamma_\alpha(1+\rho\gamma_5)\frac{\hat{p}-\hat{k}+m}{-2pk}\hat{e}u(p).
\ea

We extract the terms which correspond to the collinear kinematics of
photon emission (photon momentum is collinear to the electron one)
and the noncollinear ones. The contribution first one  contains the
large logarithm $L$, and thus include electron mass singularities.
The contribution second of one is finite in the mass electron zero
limit \ba
\sum|M^\gamma|^2=\sum|M^\gamma|^2_{coll}+\sum|M^\gamma|^2_{ncoll}.
\ea

The first term ( in agreement with the prescription of quasi-real
electron method \cite{BFK} is: \ba
\frac{1}{16}\sum|M^\gamma|^2_{coll}=(\frac{G}{\sqrt{2}}V_{ud})^2S_0(p_e(1-x),p)[\frac{1+(1-x)^2}{x(1-x)}\frac{1}{p_ek}-
\frac{m^2}{(p_ek)^2}], \ea with $x=\frac{\omega}{E}$ energy fraction
of hard photon.

The relevant contribution to the cross section is \ba
d\sigma_h=\frac{\alpha}{\pi}\int_\Delta^1 dx[
\frac{1+(1-x)^2}{2x}(L-1)+K_h] d\sigma_B(p_e(1-x),p)+
\frac{\alpha}{\pi}\ln \Delta d\sigma_B(p_e,p),\nn \\
\Delta=\frac{\Delta E}{E}\ll 1. \ea

 \section{Cross section in the leading and next to leading approximation}

Let us write down the results  obtained above in the Born
approximation and their corrections connected with emission of
virtual and soft photons as: \ba
d\sigma_B+d\sigma_{vs}=d\sigma_B[1+\frac{\alpha}{2\pi}(L-1)(2\ln\Delta+\frac{3}{2})+\nn \\
\frac{\alpha}{\pi}K_{vsc}-\frac{\alpha}{\pi}\ln \Delta]. \ea Keeping
in mind the result for the contribution of hard real photon emission
(see previous section) the total sum can be written in the form: \ba
d\sigma=\int\limits_0^1
D(x,L)d\sigma_B(p_e(1-x),p)(1+\frac{\alpha}{\pi}K)dx, \ea with the
non-singlet electron Structure function defined as (\cite{KF85}):
\ba D(x,L)=\delta(x)+\frac{\alpha}{2\pi}(L-1)P^{(1)}(x)+..., \ea and
the kernel of the evolution equation of twist-2 operator \ba
P^{(1)}(x)=\lim_{\Delta\to
0}[\delta(x)(2\ln\Delta+\frac{3}{2})+\frac{1+(1-x)^2}{x}\theta(x-\Delta)].
\ea

The smoothed form of the electron structure function is \ba
&&D(x,L)=b x^{b-1}\left[1+\frac{3}{4}b\right]-b\left(1-\frac{1}{2} x \right)+O(b^2), \\
&&b=\frac{2\alpha}{\pi}(L-1); \qquad \int_0^1 D(x,L)dx=1. \nn \ea

The so-called K-factor $K=K_{EW}+K_{vsc}+K_h$ is the smooth
function, finite in the limit of zero electron mass $m\to 0$. We put
this contribution in the form $K_{EW}=(K_{EW})_\Lambda +
(K_{EW})_f$, where
\ba
    (K_{EW})_\Lambda = \left[-\frac{1}{2} + A \right] \ln\left(\frac{M_W^2}{M^2}\right),
\ea and the part which does not depend on $M_W^2/M^2$ is included in
$(K_{EW})_f$. Numerically it is much smaller than $(K_{EW})_\Lambda$
and is not touched upon here. For details see \cite{Hollik}.

The value $K_h$ depends on experimental conditions of photon
detection and will not conditional here.

The function $K_{svc}$ is presented in the Table.

\section{Crossed processes. Discussion.}

The differential cross section of process (2)\quad
$e^+(p_e)+N(p_n)\to P(p)+\bar{\nu}(p_\nu)$ can be obtained from the
results given above by replacement $s\leftrightarrow -u$ with the
same definitions of kinematic invariants.

The cross section of the process $(\nu n),(\nu p)$ (see (2)) in the
leading logarithmical approximation (LLA) has the form \ba
\frac{d}{dy}d\sigma^i(y,s)=\int_y^1\frac{dx}{x}d\sigma^i_{B}\left(\frac{p_e}{x}\right)
D\left(\frac{y}{x},L\right),i=(\nu p), (\nu n), \ea where $E_\nu,y
E_\nu$ are energies of the initial neutrino and the final lepton.

The relevant cross sections in the Born approximation have a form
(5) the squared matrix element given in (6) and phase volume defined
(7,8).

One can be convinced in the fulfillment of the
Kinoshita-Lee-Nauenberg theorem \cite{Kinoshita:1962ur,Lee:1964is}
about cancellation of mass singularities in the fraction of final
electron differential cross sections integrated by energy: \ba
\int_0^1 dy \frac{d}{d y}d\sigma(y)=\int_0^1 dx
d\sigma_B\left(\frac{p_e}{x}\right)
\int_0^x\frac{dy}{x}D\left(\frac{y}{x},L\right)= \int_0^1dx
d\sigma_B\left(\frac{p_e}{x}\right). \ea So the dependence on $L$
disappears.

The order of value of the total  cross sections considered above
$$\sigma\sim 10\mbox{pb}~E(GeV)$$
is too small to be measured in modern experiments.

Above, we supposed to simplify the structure of hadron weak current.
This really it must be considered as a model approximation. In
reality, the induced axial and vector nucleon formfactors must be
taken into account \cite{Okun}. Nevertheless, the form cross section
(33) remains valid.

\section*{Acknowledgements}
 One of us (EAK) iacknowledges the support of INTASS grant
 05-1000008-8328.
We are grateful to G. G. Bunatian for fruitful discussions.

\appendix

\section{K-factor}

The relevant loop momentum integrals are \ba (I,I_\mu,
I_{\mu\nu},I_{kk})=\int\frac{d^4k}{i\pi^2}\frac{1,k_\mu,k_\mu k_\nu,
k^2}{(0)(1)(2)}. \ea Feynman joining denominators procedure leads to
(scalar 3-denominator integral $I$ is given above): \ba
I_\mu=-p_{e\mu}i_x+p_\mu i_{\bar{x}}; \nn \\
I_{\mu\nu}=\frac{1}{4}g_{\mu\nu}[L_\Lambda-i_p-\frac{1}{2}]-\frac{1}{2}[p_{e\mu}
p_{e\nu}I_{xx}+
p_\mu p_\nu i_{\bar{x}\bar{x}}-(p_{e\mu} p_\nu+p_{e\nu} p_\mu)I_{x\bar{x}}]; \nn \\
I_{kk}=L_\Lambda-i_p-1. \ea

The values of the contributions to the $K_{svc}=-K_V+K_S+K_C$-factor
are: \ba K_S=1+\xi_2; \qquad K_C=-\frac{9}{4}-\frac{1}{2}L_M, \ea
and \ba
K_V=-\frac{1}{2}L_M^2-3\xi_2-Li_2\left(1+\frac{M^2}{s}\right)+\frac{1}{2
S_0 a}\left[ B_1 + \frac{1}{a} L_M B_M\right], \ea where

\ba
B_1=-M^4(s+u)(1-\rho)(7+11\rho)+(s+M^2)u^2(1-\rho)^2+8s^3(1+\rho)^2+\nn\\M^2s^2(1+\rho)(-3+19\rho)-9suM^2(1-\rho^2); \nn \\
B_M=-4M^6(s+u)(1-\rho^2)+s^2u^2(1-\rho)^2+4s^4(1+\rho)^2+\nn\\u^2(s+M^2)M^2(1-\rho)^2-M^2s^2u(1-\rho)(7+5\rho)+ \nn \\
+M^2s^3(1+\rho)(-1+13\rho)-10M^4su(1-\rho^2)+M^4s^2(-7+2\rho+13\rho^2).
\ea

\begin{figure}
\begin{center}
\includegraphics[width=0.8\textwidth]{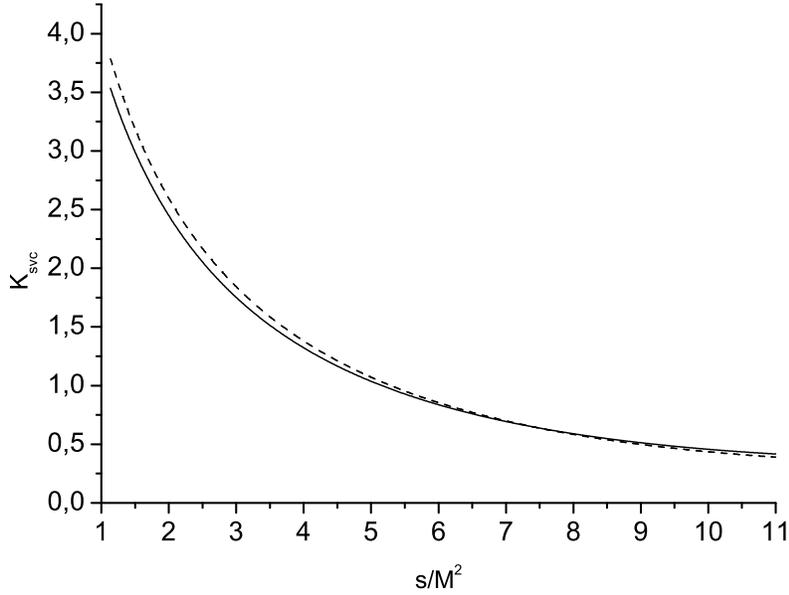}
\caption{Dependence of $K_{svc}$ on $(s/M^2)$ for $u=-s$ (dashed
line) and $u=0$ (solid line) (see Appendix).} \label{FigKSVC}
\end{center}
\end{figure}

In Fig. \ref{FigKSVC} the dependence of $K_{svc}(s,u)$ is presented
as a function of $s$ for fixed values of $u$.

\end{document}